\newif\ifOneCol 

\ifOneCol
\documentclass[journal,12pt, onecolumn, draftclsnofoot]{IEEEtran}
\else
\documentclass[[10pt, conference, table]{IEEEtran}
\fi

\IEEEoverridecommandlockouts
\usepackage{cite}
\usepackage{verbatim}
\usepackage{url}
\usepackage{grffile}
\usepackage{graphicx}
\usepackage{subcaption}
\usepackage{amsmath,amssymb,amsfonts}
\usepackage[detect-all]{siunitx}
\usepackage[hidelinks]{hyperref}
\usepackage{cleveref}
\Crefname{section}{Sec.}{Secs.}
\usepackage{tikz}
\usepackage{algorithmic}
\usepackage{epstopdf}
\usepackage[utf8]{inputenc}
\usepackage{booktabs}
\usepackage{hhline} 
\usepackage{highlight}
\usepackage{color,xcolor} 
\renewcommand{\opacity}{50}

\def\BibTeX{{\rm B\kern-.05em{\sc i\kern-.025em b}\kern-.08em
    T\kern-.1667em\lower.7ex\hbox{E}\kern-.125emX}}

\newcommand\copyrighttext{%
  \footnotesize This work is licensed under a Creative Commons Attribution 4.0 License. For more information, see \url{https://creativecommons.org/licenses/by/4.0/}}
\newcommand\copyrightnotice{%
\begin{tikzpicture}[remember picture,overlay]
\node[anchor=south,yshift=10pt] at (current page.south) {\fbox{\parbox{\dimexpr\textwidth-\fboxsep-\fboxrule\relax}{\copyrighttext}}};
\end{tikzpicture}%
}    

\usepackage[]{footmisc}

\begin{document}
\bstctlcite{IEEEexample:BSTcontrol}

\title{On the Application of Reliability Theory to Cellular Network Mobility Performance Analysis  \vspace{-0.4\baselineskip}}

\author{\IEEEauthorblockN{Subhyal Bin Iqbal$^{*\dagger}$, Behnam Khodapanah$^{*}$, Philipp Schulz$^{\dagger}$ and Gerhard P. Fettweis$^{\dagger}$ \vspace{-0.94\baselineskip}}  \\
\IEEEauthorblockA{$^{*}$ Nokia Standardization and Research Lab, Munich, Germany  \vspace{-0.94\baselineskip}}  \\
{$^{\dagger}$ Vodafone Chair for Mobile Communications Systems, Technische Universität Dresden, Germany  
 \vspace{-\baselineskip}} 
}


\markboth{IEEE Wireless Communication Letters}%
{IEEE WCL \LaTeX \ Templates}

\maketitle

%
\copyrightnotice

\vspace{-0.8\baselineskip} 

\begin{abstract}
Achieving connectivity reliability is one of the significant challenges for 5G and beyond 5G cellular networks. The present understanding of reliability in the context of mobile communication does not adequately cover the stochastic temporal aspects of the network, such as the duration and spread of packet errors that an outage session may cause. Rather, it simply confines the definition to the percentage of successful packet delivery. In this letter, we offer an elaborate modeling of the outage for a cellular mobile network by showcasing the different types of outages and their contiguity characteristic. Thereafter, using the outage metrics, we define two new key performance indicators (KPIs), namely \textit{mean outage time} and \textit{mean time between outages} as counterparts to akin KPIs that already exist in classical reliability theory, i.e., mean down time and mean time between failures. Using a system-level simulation where user mobility is a crucial component, it is shown that these newly defined KPIs can be used to quantify the reliability requirements of different user applications in cellular services. 

\end{abstract}


\vspace{-0.2\baselineskip}

\section{Introduction} \label{Section1}

One of the key objectives of fifth-generation (5G) and futuristic sixth-generation (6G) mobile networks is to improve the overall reliability in frequency range 2 (FR2) \cite{38331}. In classical terms, reliability can be described as the probability that a system in a given operational state will perform the required function for a specified time \cite{Reliabilitytheorybook}. In the more specific context of wireless communication, \textit{3GPP} has outlined reliability as the probability that a certain amount of data is successfully transmitted from the source node to the destination node within a specified delay \cite{38913}. However, there are a few limitations to this definition. Firstly, this definition does not consider the system's initial state, which is key when considering the classical definition of reliability. Secondly, it ignores its dependency on the time dimension altogether \cite{RealiabilitySoTO}. Furthermore, it is centered more toward the ultra-reliable low latency communication (URLLC) use case of 5G. To the best of our knowledge, leveraging well-accepted definitions and methods of reliability theory to wireless communication concerning
temporal aspects has been addressed only by a few researchers. In \cite{RealiabilitySoTI}, the expected reliability and mean time to first failure of a wireless mobile system are studied but the communication links are assumed as perfect. \textcolor{black}{The study considers multiple mobile hosts which roam between cells, where the sojourn time within a cell and the handover completion time are assumed to be exponentially distributed random variables}. In \cite{RealiabilitySoTO}, concepts of reliability theory are applied and extended to URLLC networks with non-mobile users, wherein the networks are modeled as being repairable and failures stem from co-channel interference. In another related work \cite{ReliabilitySOTII}, analytical expressions for time-dependent reliability key performance indicators
(KPIs) are derived for a wireless URLLC multi-connectivity scenario with Rayleigh fading, wherein a single user with no user mobility is considered. \textcolor{black}{The study in \cite{StateoftheArtLatetsReference} extends the reliability theory analysis of \cite{ReliabilitySOTII} to a multi-user static system for hybrid visible light communication-radio frequency networks. In \cite{GerhardAGVPaper}, the applications of reliability theory in terms of mission reliability are discussed, where multiple mobile autonomous guided vehicles in a campus-wide edge network setting for industrial environments are considered}.

In this letter, we model a 5G beamformed mobile network as defined in \cite{SubhyalMPUEpaper} as a repairable system, considering the enhanced mobile broadband (eMBB) use case of 5G. The system can be described as having an operational state characterized by ideal nodes (i.e., user equipment (UE), base station (BS)) and non-ideal links that are subject to radio link conditions. The intended function is defined as being in a state whereby data can be transmitted to the network, which is only possible by maintaining connectivity with the serving cell. Lastly, the specified time is characterized as the time duration in which the UE is mobile within the 5G beamformed network. This interpretation is linked closely to mobility interruption time defined in \textit{3GPP TR 38.913} \cite{38913}, which can be delineated as the shortest time duration supported by the system during which a UE cannot exchange data with the network. Mobility interruption time can alternatively be defined as outage per \textit{3GPP} nomenclature~\cite{36881}. In  \cite{38913} it is stated that the “target for mobility interruption time should be 0\,ms", and this is the only explicit time attribute in the current discussion. However, such an optimum cannot be guaranteed due to the underlying radio link conditions in 5G networks \cite{SubhyalMPUEpaper}. 

In this letter, outage as the cause of a failure or down state for a repairable system within the classical reliability theory context \cite{Reliabilitytheorybook} is modeled comprehensively for a cellular mobile network. Thereafter, we take the time dimension into account and introduce two novel KPIs based on outage modeling as counterparts to existing ones from the classical case. This is important since in mobile communication not only the probability of outage but also the duration and time spacing of outage sessions determines the quality of service (QoS) and the quality of experience (QoE). The derived KPIs are of great interest in the design of cellular networks that will cater to the different user applications within the eMBB use case and its future extensions in 6G. Conducting system-level simulations for an exemplary scenario, the mobility, outage and reliability theory KPIs are then jointly evaluated. 





\section{Network Model} \label{Section2}


The considered 5G beamformed network model includes all the major relevant functionalities applicable to the physical and the medium access layer \cite{38901, 38802, SubhyalMPUEpaper}. A list of the complete simulation parameters can be found in \cite[Sec.\,III.A]{Subhyalhandblockagepaper}. An outdoor urban-micro cellular deployment consisting of a standard hexagonal grid with seven sites, each divided into three cells, is considered, as defined in \textit{3GPP TR 38.901} as part of the evaluation standards\cite{38901}. \textcolor{black}{More details of the exact network layout can be found in Fig. 4 of \cite{Subhyalhandblockagepaper}}. Although the hexagonal grid model might differ from realistic deployments, this simplification offers an equitable abstraction to existing 5G and beyond 5G networks. The carrier frequency is \SI{28} {GHz}, and the inter-site distance is \SI{200}{m}.    The transmitter (Tx)-side beamforming model is based on \cite{Umurchannelmodelpaper}, where a 12-beam grid configuration is considered. $K_b=4$\,beams are simultaneously scheduled for all cells in the network. Beams $b \in \{1,\ldots,8\}$ have smaller beamwidth and higher beamforming gain and cover regions further apart from the BS. Beams  $b \in \{9,\ldots,12\}$ have larger beamwidth and relatively smaller beamforming gain and cover regions closer to the BS. 
\textcolor{black}{A wrap-around is considered, meaning that the hexagonal grid is repeated around the original layout in six replicas in order to avoid boundary effects concerning interference}. Further details of the channel model relating to shadow and fast fading can be found in \cite[Sec.\,III]{SubhyalMPUEpaper}.



\begin{figure}[!t]
\textit{\centering
    \ifOneCol
        \includegraphics[width=12cm]{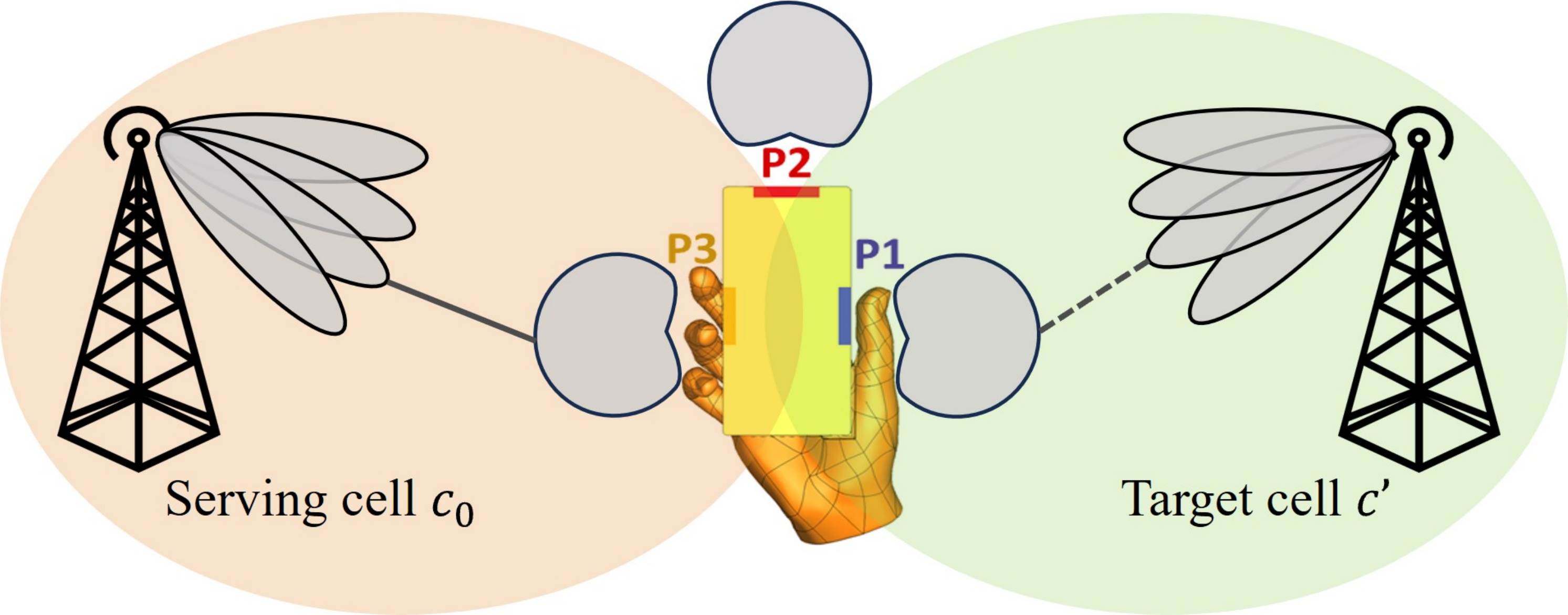}
    \else
    \includegraphics[width = \columnwidth]{Fig1.5-updated.pdf}
    \fi
\vspace{-\baselineskip}
\caption{\textcolor{black}{An illustration of an MPUE in the \textit{edge} design with three directional panels, shown here for an exemplary 5G beamformed network.}} 
\label{fig:Fig1}}
\vspace{-\baselineskip} 
\end{figure}

At the start of the simulation, $N_\textrm{UE}=420$\,UEs are dropped randomly in the aforementioned network deployment scenario. The multi-panel UE (MPUE) architecture assumes an \textit{edge} design for the UEs with three directional panels mounted on the top, left, and right edges of the UEs \cite{SubhyalMPUEpaper}. \textcolor{black}{A schematic diagram for the MPUE is shown in Fig.\,\ref{fig:Fig1}}. The UEs follow a 2D uniform distribution over the network, moving at constant velocities along straight lines into random directions \cite[Table 7.8-5]{38901}. Following \textit{3GPP}, the signal measurement scheme that we consider in this study is MPUE-A3, where it is assumed that the UE can measure the reference signal received power (RSRP) values from the serving cell $c_0$ and neighboring cells by simultaneously activating all of its three panels \cite{SubhyalMPUEpaper}. The simulation run-time is $T_\textrm{sim}$ = \SI{30}{s}.


The handover model considered in this letter is conditional handover (CHO), introduced in \textit{3GPP Release 16} \cite{CHOTechRep}. In CHO the handover is prepared early, when the UE-serving cell link is sufficiently strong, but the actual handover execution happens later when the UE-target cell link is sufficient \cite[Sec.\,II.B]{Subhyalhandblockagepaper}. Within the network, each UE is assumed to be capable of measuring the raw RSRP values $P_{c,b}^\textrm{RSRP}(n)$ at a discrete time instant $n$ from each Tx beam $b \in B$ of cell $c \in C$, using the synchronization signal block (SSB) bursts that are periodically transmitted by the BS. At the UE side, layer 1 (L1) and L3 filtering are then sequentially applied to the raw RSRPs to counter the effects of fast fading and measurement errors. The L3 measurements are an indicator of the average downlink signal strength for a link between a UE and cell~$c$. They are used in the conditional handover for both handover preparation and execution. A more detailed explanation of the L1 and L3 filtering procedures can be found in \cite[Sec.\,II.A]{SubhyalMPUEpaper}. 


The beam management procedure considered in this letter is based on \cite{38802}, whereby within each of the twenty-one serving cells, a mobile UE may switch to a different serving beam $b_0$ based on L2 measurements. In the event of a beam failure detection (BFD) due to the radio link quality metric (RLQ) falling below a certain signal-to-noise-plus-interference ratio (SINR) ratio $\gamma_\textrm{out}$  = \SI{-8}{dB}, the UE initiates a beam failure recovery (BFR) process where it tries to recover to a stronger beam of the same serving cell \cite[Sec.\,II.B]{SubhyalMPUEpaper}. If the BFR process is also unsuccessful, the UE declares a radio link failure (RLF) and tries to re-establish the connection to the same serving cell or a different neighboring cell using random access. 

The link quality between a UE and the serving cell $c_0$ through its serving beam $b_0$ is specified through the average downlink SINR $\gamma_{c_0,b_0}$. The average downlink SINR is a discrete-time value taken as the expected value of the instantaneous downlink SINR, which is modeled as a random variable \cite{AmanatdownlinkSINR}. As will be seen later in \Cref{Subection3.2}, the average downlink SINR has a key role in modeling different mobility events that lead to outages.

\section{Reliability Theory System Model} \label{Section3}
In this section, the KPIs pertaining to classical reliability theory are discussed first. Thereafter, their mobility counterparts are discussed. 

\subsection{Reliability Theory in the Classical Sense} 
\label{Subection3.1}

Reliability theory is a branch of systems engineering concerned with determining the dependability and soundness of a system. This is accomplished by determining the probability that a system will perform the intended function for a prescribed period under the given environmental conditions \cite{Reliabilitytheorybook}. A system can be either repairable or non-repairable. In a repairable system, where failures appear and successful repairs are performed, “up” and “down” states occur. 
The corresponding up and down times describe the cumulative time spent in the up and down states, respectively. A handful of KPIs have been defined to ascertain the performance of a repairable system  \cite{Reliabilitytheorybook}. This letter discusses two of the most widely used ones, each explained below.

\textit{Mean time between failures (MTBF)}: the average time duration between consecutive transitions from an up state to a down state. It can be described as the average time between failures in a system. 

\textit{Mean down time (MDT)}: the mean time from when the system enters a down state until it has been repaired and transitions again to an up state. It can be described as the average time duration of a system failure.  

\vspace{-0.22\baselineskip}

\subsection{Reliability Theory Applied to Mobility} \label{Subection3.2}

 In a communication network, a prerequisite for two nodes to communicate with each other is to have at least one operating path connecting them \cite{Realability2nodepaper}. This implies that all the nodes should be in an operational state, i.e., they function as intended, and all the links should be functional, i.e., they allow communication between the nodes. Only then can an analysis of reliability theory be applied to the network. As such, in this letter, a 5G beamformed network is modeled to represent a system where a mobile UE maintains constant connectivity with the network and can transmit and receive data. The operational state considers that all the UEs and cells in the network are always functioning and experience zero device-related or network-related down time, respectively. It also considers that the UE-serving cell $c_0$ link is the existent functional link through which the UE maintains connectivity to the network. The system's  susceptibility to failures arises from  outage, which is the time duration during which the UE cannot communicate with the network \cite{36881}. An outage can occur due to adverse radio link conditions or delays caused by signaling during handovers \cite[Sec.\,V.B]{SubhyalMPUEpaper}. In the case an outage is experienced, the system can recover and go back to a regular data transmission/reception state, i.e., it is repairable. This modeling approach is more in line with the classical definition of reliability theory defined in \cite{Reliabilitytheorybook} than the \textit{3GPP} definition in \textit{TR 38.913} \cite{38913} since it takes into account the initial operational state of the system. Moreover, it also now considers the time-based dependability attributes for the eMBB use case of 5G by introducing two new KPIs.
 
 In this letter, the outage is elaborately modeled by categorizing it into several distinct components.

\subsubsection{Successful handover outage} The outage when the UE is in the process of executing a handover from the serving cell $c_0$ to the target cell $c^{\prime}$. As per \textit{3GPP Release 16} \cite{HOInterruptionTimeReductionTechRep}, this outage is modeled as a fixed duration of $T_{\mathrm{HO}} = \SI{55}{ms}$. In line with \textit{3GPP} \cite{38331}, the outage due to multiple RACH attempts $N_\textrm{Att}$ that lead to a successful handover is also considered.

 \subsubsection{Handover failure outage} The outage when the handover to the target cell fails due to adverse radio link conditions  \cite[Sec.\,IV]{SubhyalMPUEpaper}. This outage includes both the duration of both the handover failure timer $T_{\mathrm{HOF}} = \SI{200}{ms}$ and the re-establishment time $T_{\mathrm{res}} = \SI{180}{ms}$ it takes for the UE to reconnect to the same cell or to a different target cell \cite{38331}. 

 \subsubsection{BFR outage} The outage due to the BFR process that follows BFD, as discussed in \Cref{Section2}. This includes the outage in the event that the BFR is both successful or unsuccessful.

 \subsubsection{RLF outage} The outage due to an RLF, which occurs when the UE fails to recover to another beam within the serving cell and the BFR process fails, as discussed earlier in \Cref{Section2}. In this case the UE would reconnect to a different cell and the RLF outage duration is $T_{\mathrm{res}} = \SI{180}{ms}$ \cite{38331}.
 
The sum of handover failure outage and RLF outage is denoted as \textit{mobility failure outage}.

 \subsubsection{SINR degradation outage} The outage when a  UE experiences adverse radio conditions due to the SINR falling below the threshold $\gamma_\mathrm{out}$ = \SI{-8}{dB} \cite{SubhyalMPUEpaper}. Excludes the cases wherein the UE is executing a handover, is in the process of re-establishment, or in BFR. Herein, the average downlink SINR, discussed in \Cref{Section2}, is used as the reference SINR.

Outage of any type is denoted in terms of a percentage as
%
\begin{equation}
\label{Eq1} 
\textrm{Outage} \ (\%) = \frac{\sum_{u}{\textrm{Outage duration of UE}} \ u} {N_\mathrm{UE} \ \cdot \ T_\textrm{sim}} \ \cdot \ 100. 
\end{equation}
%

 An essential contribution of this letter is to identify that the aforementioned outage components can occur independently or contiguously, with one or more outage components occurring successively. The latter case leads to a single long outage session. For example, on cell boundaries, it is very common that a UE may experience an outage due to SINR degradation, which can be immediately followed by the UE also experiencing BFD, leading to BFR, which is eventually successful when the UE reconnects to another beam of the serving cell. However, during the BFR procedure the conditional handover execution condition could also be triggered \cite[Sec.\,II.B]{Subhyalhandblockagepaper}, leading to a handover. This is shown in Fig.\, \ref{fig:Fig3}. Identifying such contiguous outage sessions is necessary to correctly consider the system's temporal characteristics and accurately determine reliability metrics related to mobility. \Cref{Table1} shows a detailed description of the aforementioned outage types.

 
\begin{figure}[!t]
\textit{\centering
    \ifOneCol
        \includegraphics[width=10cm]{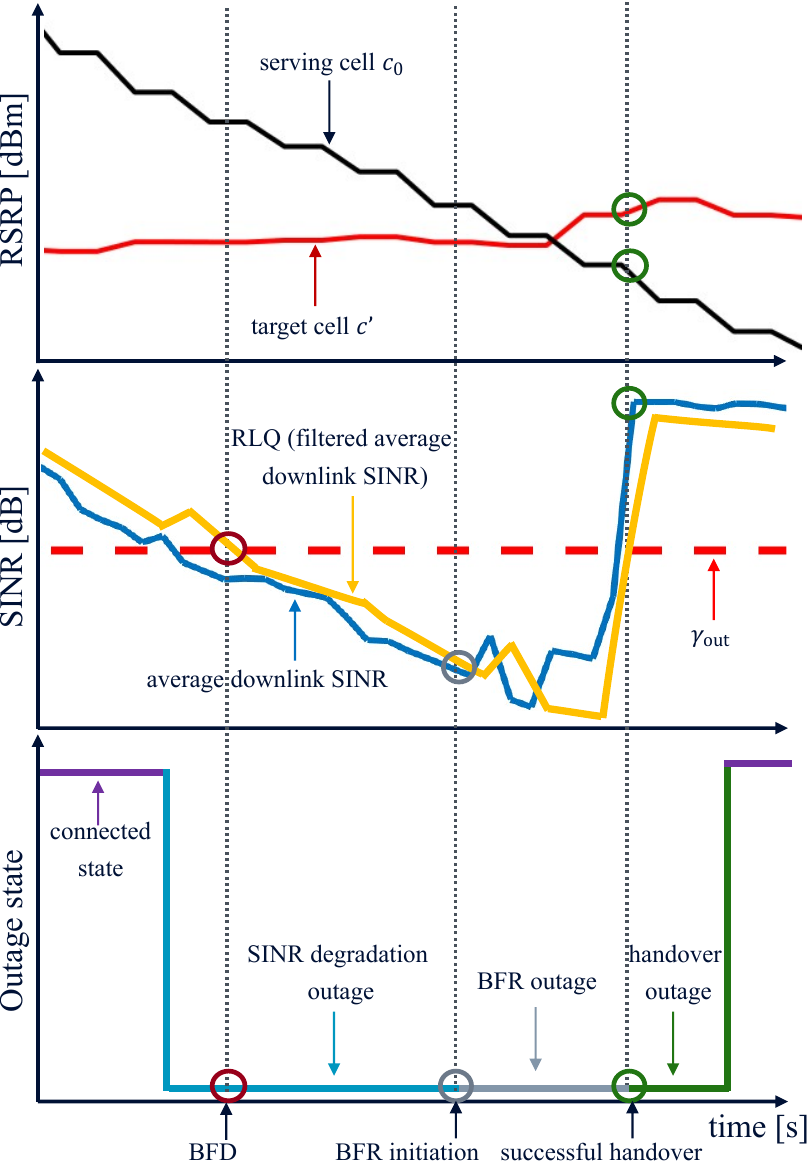}
    \else
        \includegraphics[width = 0.96\columnwidth]{Fig8-new_updated-III.pdf}
    \fi
\vspace{-2\baselineskip}
\label{fig:Fig3}
\vspace{2.2\baselineskip}  
\caption{A depiction of a contiguous outage session in the 5G network model.}
\label{fig:Fig3}}
\vspace{-0.7\baselineskip} 
\end{figure}

   \ifOneCol
    \begin{table}[!t]
    \else
    \begin{table}[!t]
    \fi

\begin{center}
\caption{\vphantom{\rule[0.0in]{2pt}{\baselineskip}}%
Outage types and their contiguity characteristic}
 \vspace{-2pt}
\label{Table1}
\begin{tabular}{ll}
\toprule 
\bfseries Outage type  & \bfseries  Contiguity characteristic  \\
\midrule 
Successful handover & May be preceded by outage due to successful  \\
outage &  BFR attempts, where both cases can be preceded   \\
&   by outage due to SINR degradation \\
\midrule 
Handover failure & May be preceded by outage due to successful  \\
outage &   BFR attempts, where both cases can be preceded   \\
&   by outage due to SINR degradation \\
\midrule 
RLF outage & Always preceded by outage due to BFR failure,  \\ 
 &  which can be preceded by outage due  to SINR  \\
&  degradation \\
\midrule 
BFR success  & Standalone or preceded by outage due to SINR \\
outage &  degradation \\
\midrule 
Standalone SINR  & Standalone outage by definition \\
degradation outage & \\
\bottomrule
\end{tabular}
\end{center}
 \vspace{-16pt}
    \ifOneCol
    \end{table}
    \else
    \end{table}
    \fi

Herein, two novel and promising KPIs are defined as counterparts to MTBF and MDT. They are an effort to link reliability theory with  the time dimension for 5G networks, where user mobility is a key factor to the analysis. By doing so, we step beyond  the rather rudimentary \textit{3GPP} definition of defining reliability as the percentage of successful packet delivery \cite{38913}. The KPIs are explained below.

\textit{Mean time between outages (MTBO)}: the average time duration between individual outage sessions, defined as
\vspace{-0.3\baselineskip}  
\begin{equation}
\label{Eq2} 
\textrm{MTBO} = \frac{N_\mathrm{UE}  \cdot T_\textrm{sim} - \sum_{s \in \mathcal S} t_s^\textrm{O}} {|\mathcal S|},
\end{equation}
where $\mathcal S = \bigcup_{u \in \mathcal U} \mathcal S_u$, with $\mathcal S_u$ being the set containing the individual outage sessions of each UE $u \in U$ and $t_s^\textrm{O}$ being the duration of session $s$. The set cardinality is denoted by $|\cdot|$.

\textcolor{black}{The set $\mathcal S_u$ is important since empirical distributions can then be studied with respect to  percentiles to infer if performance guarantees of eMBB applications can be met}.

\textit{Mean outage time (MOT)}: the average time duration of an outage session, defined as 
\vspace{-0.3\baselineskip}  
\begin{equation}
\label{Eq3} 
\textrm{MOT} = \frac{\sum_{s \in \mathcal S}t_s^\textrm{O}}   {|\mathcal S|}.
\end{equation}
Both KPIs are expressed in seconds (s).

\section{Performance Evaluation} \label{Section4}
In this section, the novel KPIs of MTBO and MOT are jointly evaluated with the mobility and outage KPIs. The considered exemplary scenario is described in \cite[Sec.\,IV.B]{Subhyalhandblockagepaper}, where the MPUEs are assumed either in the free space (FS) setting or with hand blockage (HB) on panel 1 (P1) and P3.


\textcolor{black}{In \Cref{Table2}, the two key mobility KPIs of successful handovers and mobility failures are shown, normalized to the number of UEs $N_\mathrm{UE}$ in the network and to time (expressed as \SI{}{KPI/UE/min}}).  It can be seen in \Cref{Table2} that as the UE speed increases, both successful handovers and mobility failures increase. For example, for the FS case, as the UE speed increases from \SI{60}{km/h} to \SI{120}{km/h}, the number of successful handovers increases almost two-fold (from \SI{19.30}{} to \SI{31.29}{KPI/UE/min}). The mobility failures also increase correspondingly by almost 5x (from \SI{0.48}{} to \SI{2.45}{KPI/UE/min}). Since the UE also traverses more cell boundaries at higher speeds, the probability of both handovers and mobility failures increases. This is also represented by the increase in the corresponding outage components of successful handovers and mobility failures in \Cref{Table3}. Compared to \SI{60}{km/h}, at \SI{120}{km/h} mobility becomes more challenging because of greater temporal variations in the signal RSRPs due to dominant fast fading. This is the other significant reason why more  mobility failures are observed at higher speeds. The temporal variations also lead to adverse radio link conditions and hence the outage due to SINR degradation and BFR also increases in \Cref{Table3}. As a result, the total outage is also seen to increase.

\begin{figure}[!t]
\textit{\centering
    \ifOneCol
        \includegraphics[width=15cm]{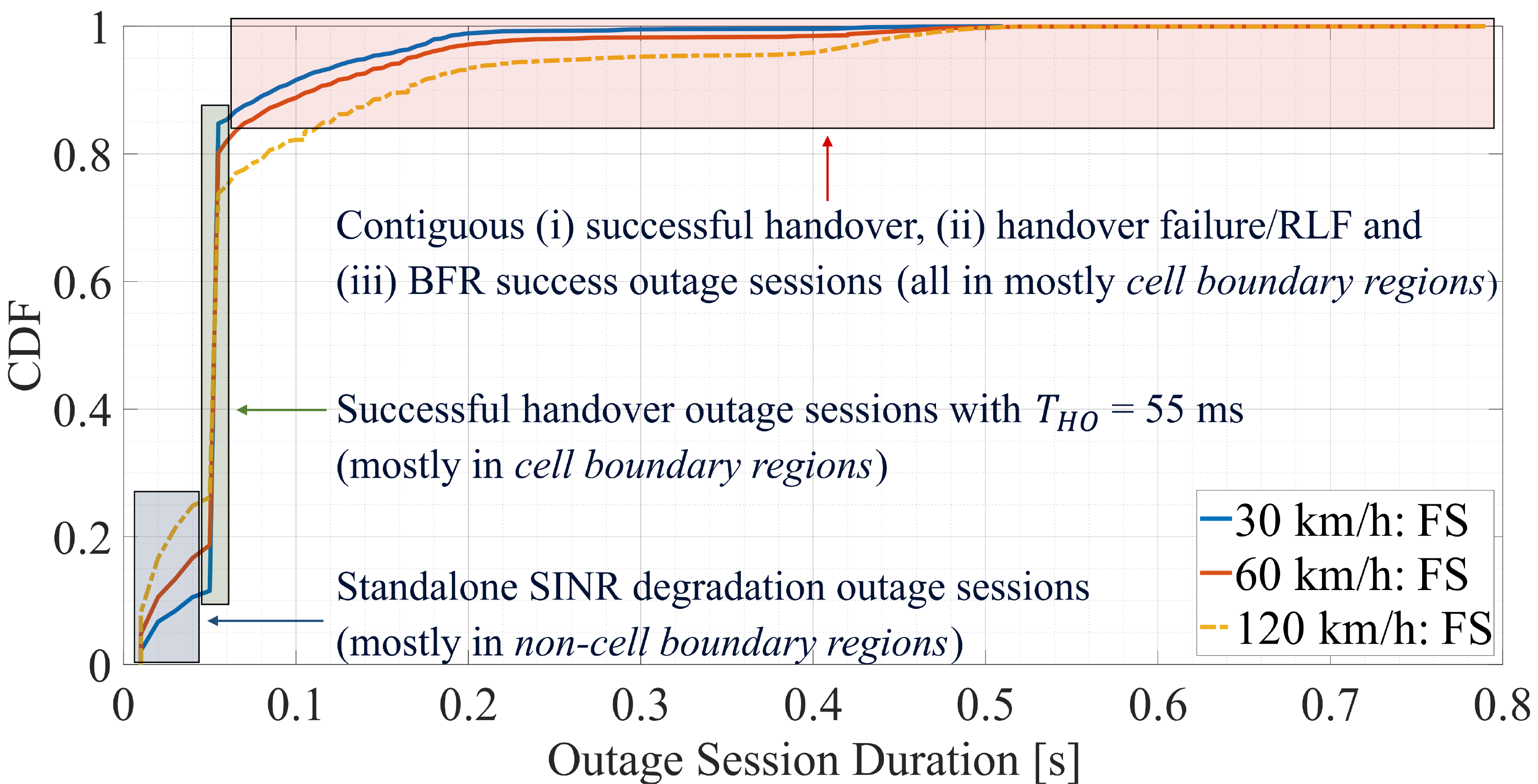}
    \else
    \includegraphics[width = 0.99\columnwidth]{Fig12.pdf}
    \fi
\vspace{-1.1\baselineskip}
\caption{\textcolor{black}{The empirical CDF of the total outage session duration, shown here for the free space (FS) case. The three distinct outage regions and their sources are also depicted.}} 
\label{fig:Fig4}}
\vspace{-0.9\baselineskip} 
\end{figure}

This analysis of the mobility and outage KPIs is nicely captured in \Cref{Table4}, where the MOT increases by around 20\% (\SI{68}{ms} to \SI{81}{ms}) when \SI{60}{km/h} is compared with \SI{120}{km/h}. The increase is observed because of the significant increase in outage due to mobility failures, where the re-establishment time alone is \SI{180}{ms} (as discussed in \Cref{Subection3.2}). On the other hand, MTBO is almost halved (from \SI{2.16}{} to \SI{1.14}{s}) because the total number of outage sessions due to successful handovers, mobility failures, SINR degradation, and BFR increase. This translates into outage sessions occurring more frequently and hence less mean time between the individual sessions. The inference here is that reliability theory KPIs can be sufficient to give the whole mobility performance outlook that previously had to be understood through jointly analyzing the mobility and outage KPIs. \textcolor{black}{The MOT of successful handovers and mobility failures is also shown}. \textcolor{black}{An empirical cumulative distribution function (CDF) of the individual contiguous outage sessions for the FS case is shown in Fig.\, \ref{fig:Fig4}, where it is seen that the maximum outage session duration observed is \SI{790}{ms}, for both the \SI{60}{km/h} and \SI{120}{km/h} cases.} \textcolor{black}{From a practical point of view, these KPI values can be understood through cloud gaming and extended reality (XR), both popular user applications of eMBB}. An average outage duration of \SI{68}{ms} at an interval of \SI{2.17}{s} compared to an average outage duration of \SI{81}{ms} at an interval of \SI{1.14}{s} could lead to an entirely different QoE for cloud gaming, where the network time delay requirements usually lie between \SI{100}{ms}-\SI{150}{ms} \cite{QoEReference}. As such, an outage session coupled with latency in the access and core network are more likely to not satisfy the QoE requirements for  UEs traveling at higher  than at lower speeds. \textcolor{black}{For XR, the time delay requirements lie between \SI{10}{ms}-\SI{30}{ms} \cite{NokiaXRPaper}, depending on the traffic type, e.g., video streaming or motion and control}. \textcolor{black}{For applications involving buffering such as video streaming, MOT values smaller than the buffer size and MTBO values larger than the payload to fill the buffer size/throughput  may have negligible impact on the QoS and QoE.}

\begin{table}[!t]
\begin{center}
\caption{\textcolor{black}{Mobility KPIs (in KPI/UE/min)}}
\label{Table2}
\begin{tabular}{| c | c | c |}
\hline
\bfseries Case & \bfseries Succ. Handovers  & \bfseries Mob. Failures \\
\hline
\bfseries 30 km/h: FS  &  \gradientcell{10.60}{10.10}{31.29}{white}{green}{25} & \gradientcell{0.07}{0.07}{3.15}{white}{red}{20}\\
\bfseries 30km/h: HB  &  \gradientcell{10.10}{10.10}{31.29}{white}{green}{15}  & \gradientcell{0.14}{0.07}{3.15}{white}{red}{30}\\
\bfseries 60 km/h: FS  &  \gradientcell{19.30}{10.10}{31.29}{white}{green}{40}  & \gradientcell{0.48}{0.07}{3.15}{white}{red}{55}\\
\bfseries 60km/h: HB  &  \gradientcell{18.17}{10.10}{31.29}{white}{green}{30}  & \gradientcell{0.59}{0.07}{3.15}{white}{red}{87}\\
\bfseries 120 km/h: FS  & \gradientcell{31.29}{10.10}{31.29}{white}{green}{66}  & \gradientcell{2.45}{0.07}{3.15}{white}{red}{82}\\
\bfseries 120 km/h: HB  &  \gradientcell{29.01}{10.10}{31.29}{white}{green}{42}  & \gradientcell{3.15}{0.07}{3.15}{white}{red}{93}\\
\hline
\end{tabular}
\end{center}
\vspace{-8pt}
\end{table}

\begin{table}[!t]
\begin{center}
\caption{\textcolor{black}{Outage KPIs (in \%)}}
 \vspace{-2pt}
\label{Table3}
\begin{tabular}{|c|c|c|c|c|c|}
\hline
\bfseries Case & \bfseries Total  & \bfseries Succ. & \bfseries Mob. & \bfseries SINR  & \bfseries BFR\\
 \bfseries & \bfseries  Outage  & \bfseries Handov.  & \bfseries Failures & \bfseries Degr. & \bfseries \\
\hline
\bfseries 30 km/h: FS  & \gradientcell{1.39}{0.02}{7.60}{white}{red}{22}  & \gradientcell{0.97}{0.02}{7.60}{white}{red}{20} & \gradientcell{0.02}{0.02}{7.60}{white}{red}{10} & \gradientcell{0.36}{0.02}{7.60}{white}{red}{32} &  \gradientcell{0.03}{0.02}{7.60}{white}{red}{18}\\
\bfseries 30 km/h: HB  &  \gradientcell{1.42}{0.02}{7.60}{white}{red}{29}  & \gradientcell{0.92}{0.02}{7.60}{white}{red}{17} & \gradientcell{0.05}{0.02}{7.60}{white}{red}{10} & \gradientcell{0.41}{0.02}{7.60}{white}{red}{44}& \gradientcell{0.04}{0.02}{7.60}{white}{red}{22}\\
\bfseries 60 km/h: FS  &  \gradientcell{3.02}{0.02}{7.60}{white}{red}{48} & \gradientcell{1.75}{0.02}{7.60}{white}{red}{27} & \gradientcell{0.16}{0.02}{7.60}{white}{red}{20} & \gradientcell{1.00}{0.02}{7.60}{white}{red}{52} & \gradientcell{0.11}{0.02}{7.60}{white}{red}{38}\\
\bfseries 60 km/h: HB  &  \gradientcell{3.19}{0.02}{7.60}{white}{red}{58}  & \gradientcell{1.65}{0.02}{7.60}{white}{red}{21} & \gradientcell{0.20}{0.02}{7.60}{white}{red}{26} & \gradientcell{1.21}{0.02}{7.60}{white}{red}{62}  & \gradientcell{0.13}{0.02}{7.60}{white}{red}{42}\\
\bfseries 120 km/h: FS  &  \gradientcell{6.58}{0.02}{7.60}{white}{red}{77}  & \gradientcell{2.82}{0.02}{7.60}{white}{red}{57} & \gradientcell{0.84}{0.02}{7.60}{white}{red}{46} & \gradientcell{2.48}{0.02}{7.60}{white}{red}{48} & \gradientcell{0.44}{0.02}{7.60}{white}{red}{56}\\
\bfseries 120 km/h: HB  &  \gradientcell{7.60}{0.02}{7.60}{white}{red}{97}  & \gradientcell{2.63}{0.02}{7.60}{white}{red}{40} & \gradientcell{1.07}{0.02}{7.60}{white}{red}{54} & \gradientcell{3.30}{0.02}{7.60}{white}{red}{68} & \gradientcell{0.60}{0.02}{7.60}{white}{red}{70}\\
\hline
\end{tabular}
\end{center}
\vspace{-7pt}

\end{table}

    %
\begin{table}[!t]
\begin{center}
\caption{\textcolor{black}{Reliability Theory KPIs (in seconds)}}
 \vspace{-2pt}
\label{Table4}
\begin{tabular}{| c | c | c | c | c |}
\hline
\bfseries Case & \bfseries MOT  & \bfseries MTBO & \bfseries MOT of  & \bfseries MOT of\\
 \bfseries & \bfseries    & \bfseries  & \bfseries Succ. Handov. & \bfseries Mob. Fail. \\
\hline
\bfseries 30 km/h: FS  &  \gradientcell{0.063}{0.063}{0.092}{white}{red}{2} & \gradientcell{4.45}{1.11}{4.72}{white}{green}{40} & \gradientcell{0.060}{0.060}{0.082}{white}{red}{3} & \gradientcell{0.420}{0.420}{0.445}{white}{red}{1}\\
\hline
\bfseries 30 km/h: HB  & \gradientcell{0.068}{0.063}{0.092}{white}{red}{5}  & \gradientcell{4.72}{1.11}{4.72}{white}{green}{68} & \gradientcell{0.064}{0.060}{0.082}{white}{red}{6}  & \gradientcell{0.431}{0.420}{0.445}{white}{red}{3} \\
\hline
\bfseries 60 km/h: FS  &  \gradientcell{0.068}{0.063}{0.092}{white}{red}{5}  & \gradientcell{2.16}{1.11}{4.72}{white}{green}{37} & \gradientcell{0.065}{0.060}{0.082}{white}{red}{11}  & \gradientcell{0.445}{0.420}{0.445}{white}{red}{5} \\
\hline
\bfseries 60 km/h: HB  &  \gradientcell{0.074}{0.063}{0.092}{white}{red}{12}  & \gradientcell{2.22}{1.11}{4.72}{white}{green}{45} & \gradientcell{0.072}{0.060}{0.082}{white}{red}{15} & \gradientcell{0.445}{0.420}{0.445}{white}{red}{5}\\
\hline
\bfseries 120 km/h: FS  &   \gradientcell{0.081}{0.063}{0.092}{white}{red}{18} & \gradientcell{1.14}{1.11}{4.72}{white}{green}{20}  & \gradientcell{0.073}{0.060}{0.082}{white}{red}{19} & \gradientcell{0.431}{0.420}{0.445}{white}{red}{3}\\
\hline
\bfseries 120 km/h: HB  &  \gradientcell{0.092}{0.063}{0.092}{white}{red}{28} & \gradientcell{1.11}{1.11}{4.72}{white}{green}{16} & \gradientcell{0.082}{0.060}{0.082}{white}{red}{23} & \gradientcell{0.435}{0.420}{0.445}{white}{red}{4}\\
\hline
\end{tabular}
\end{center}
\vspace{-6pt}
\end{table}

It is also interesting to compare the impact of hand blockage on mobility performance. If the \SI{60}{km/h} case is considered, it is seen in \Cref{Table2} that  successful handovers decrease while mobility failures increase. This is due to the hand blockage effect studied in \cite{Subhyalhandblockagepaper}. Due to hand blockage, the individual outages due to SINR degradation and BFR also increase, as shown in \Cref{Table3}. This is translated into an increase of almost 10\% in the MOT value. On the contrary, the MTBO value remains almost constant. This is because the increase in the number of outage sessions due to mobility failures, standalone SINR degradation, and BFR is offset by the decrease in the number of outage sessions due to successful~handovers. 


\section{Conclusion} \label{Section4}
In this letter, we have introduced two novel KPIs for cellular mobile networks, namely MOT and MTBO. These KPIs are inspired by analogous KPIs used in classical reliability theory to determine the reliability of repairable systems. Adopting these KPIs can help propel the discussion with respect to user QoS and QoE to the next stage for 5G networks and beyond. This is because they link reliability with time, an aspect that \textit{3GPP} has so far ignored \cite{38913}. By considering a system-level simulation for an 5G network with a high mobility scenario with elaborate outage modeling, it is demonstrated that the novel KPIs offer a new dimension into understanding the mobility performance of the system. Moreover, it is also shown that the calculated values of these KPIs can be compared with the reliability metrics of practical eMBB applications such as cloud gaming and XR, where MTBO is also an important metric that should be considered to properly quantify the performance of such applications. It is observed that a difference of tens of \SI{}{milliseconds} in MOT and a few \SI{}{seconds} in the MTBO can lead to different QoS and QoE in such applications, where latency requirements can be very stringent.

\bibliographystyle{IEEEtran}
\bibliography{references}

\end{document}